\documentclass[12pt]{article}
\usepackage{graphicx}
\usepackage{psfig}
\usepackage{natbib}
\usepackage{amsfonts}
\usepackage{amssymb}
\usepackage{amsmath}
\usepackage{indentfirst}

\begin{document}
\title{Cosmic Rays and Neutrinos from GRBs: Predictions versus 
Acceleration Modeling}
\vskip 20pt
\author{\bf by D. Gialis$^{1}$ and G. Pelletier$^{1,2}$ \\
{\normalsize 1\ Laboratoire d'Astrophysique de Grenoble} \\
{\normalsize 2\ Institut Universitaire de France}}
\maketitle
\vskip 12pt
\begin{abstract}
The paper is devoted to the analysis of Fermi acceleration of protons
in GRBs and its neutrino signature. We have compared the consequences
of Bohm scaling and those
of a Kolmogorov scaling, the latter being more reliable. The
predictions about the energy limitation of UHE-protons by the various
losses and the neutrino
emissions turn out to be very sensitive to these scalings. We
consider Kolmogorov scaling as the most realistic and predict a
reasonable pp-neutrino emission around 100 GeV at the end of the
radiative stage of the fireball expansion, for a large number of GRBs
pending on their baryonic load. A second p$\gamma$-neutrino
emission is expected with the acceleration of protons in the radiation
free stage, but with a synchrotron loss limitation immediately
followed by a severe expansion loss limitation. According to the Kolmogorov
scaling, the protons could not reach the UHE-range. Anyway the large
possibility of a two component neutrino emission would be an
interesting clue of cosmic ray physics.
\end{abstract}

Keywords: GRBs, Cosmic Rays, Neutrinos, Fermi processes

\section{Introduction}
Gamma Ray Bursts (hereafter GRBs) are unique high energy phenomena in
astrophysics because of their possibility to manifest all the
interesting "astroparticles" processes, such as generation of high energy
gamma rays, ultra high energy cosmic rays, high energy neutrinos and
gravitational waves (see \citet{Dermer01} and \citet{Meszaros02}). The "fireball" model \citep{ReesMeszaros92} 
has been
successful in explaining the afterglow stage, and its more elaborated
form with the addition of internal shocks (\citet{ReesMeszaros94}, \citet{PacXu94}) during
the "free" expansion has been successful
in explaining the light curve and gamma spectra (\citet{Fishman95}, \citet{Beloborodov00}).
Because of the highly relativistic dynamics necessarly involved in the
GRB phenomenon, the generation of very high energy particles is
expected through strong shocks and strong magnetic perturbations.
Indeed GRB population could be a main source of UHE-cosmic rays which
could be generated by the external shocks \citep{Vietri95} or by the
internal shocks \citep{Waxman95}, and a flux of neutrinos produced by the
collisions of the UHE-cosmic rays with the GRB gamma photon is
reasonably expected \citep{WaxBac99}.

In this paper, our intend is to look at the sensitivity of the
predictions of cosmic ray and neutrino generation to the description
of the Fermi acceleration process together with opacity effects for
protons. We have analysed the consequences
of two assumptions, one is the so called "Bohm scaling" assumption,
the other is what we called "Kolmogorov scaling". The former consists
in the statement that the Fermi acceleration time is proportional to
the Larmor time of the accelerated particle, with a constant
proportionality factor larger than unity (often chosen between 1 and
10 in the litterature). The latter consists in taking into account
that the ratio between these two characteristic times depends on the
rigidity of the particle through a law governed by the turbulence
spectrum, as is confirmed by numerical works \citep{Casse01}
, where no Bohm
scaling has been found. In
astrophysical media such as the solar wind, the interstellar medium,
the turbulence spectrum is likely consistent with the Kolmogorov law.

The Bohm scaling is very convenient to make a first investigation of
the high energy physics performances of the objects, and our prejudices
are often grounded on this. However, in this paper, we will show that
this scaling leads to unrealistic results. The analysis based on the
Kolmogorov scaling is more reliable, and it turns out that it leads
to very different conclusions, as we will present with details in the
paper. The analysis is developed both in the radiation free stage and
in the radiative stage. We especially emphasise the regime when the 
fireball is opaque to pp-collisions during the beginning of the internal
shocks stage because
of its baryonic load, which should occur for a large fraction of GRBs
population. We make unusual predictions about the performance of
cosmic ray generation, and about neutrino emission that could have
two components, namely a non-thermal pp-neutrino emission and a
p$\gamma$-neutrino emission.

The paper is organised as follows. In section 2, we briefly present
the dynamical description we need for the estimation of the opacity
effects, not only for photons but also for protons. In section 3, we
analyse the consequence of the fast acceleration regime governed by
the Bohm scaling. In section 4, we present the properties of the more
progressive acceleration process governed by the Kolmogorov scaling
and predict a reasonable non-thermal pp-neutrino emission with its
spectrum. We investigate, more briefly in section 5, the radiation free
stage, where
UHE-cosmic rays are expected to be accelerated and p$\gamma$-neutrinos
generated. We end the paper with a discussion that summaries our
conclusions about cosmic ray and neutrino generations and the
sensitivity of the predictions to the acceleration model.

\section{Preliminary considerations}

\subsection{Dynamics of the fireball}

In this subsection, we summarize all the results we need for this
paper that describe the expansion of the fireball \citep{Meszaros93}. 
The wind flow is considered 
to be a set of discrete shells which are
successively emitted with an energy $E_{s}=E/N_{s}$, where $N_{s}$ is
the total number of shells. The duration, $t_{w}$, of this wind flow
provides with an interval of shell number, namely, $1 \leq N_{s} \leq
c\, t_{w}/r_{0}$ where $r_{0}$ is the size of the central object.\\

At the very beginning of the expansion of a shell, the pressure is
supposed to be dominated by the radiative pressure.
The temperature, $T$, of the plasma which is mainly composed by
electron-positron pairs is equal to the photon temperature.
Considering a shell is initially spherical with a radius, $r_{0}$,
and with an energy emitted in $\gamma$-rays equal to $E_{s}$, we have
\begin{eqnarray*}
\frac{E_{s}}{(4/3)\pi\, r_{0}^{3}}=a\, T^{4}\, ,
\end{eqnarray*}
where $a$ is the Stefan constant. Thus, the plasma temperature is
\begin{eqnarray*}
T=\left( \frac{3\,E_{s}}{4\pi\, a\, r_{0}^{3}} \right)^{1/4}\,.
\end{eqnarray*}

\hspace{-0.6cm}For an energy, $E$, in the neighborhood of $10^{51}$
ergs, a number of shells, $N_{s}$, of 20 and a radius, $r_{0}$, of
$10^{7}$ cm, the temperature is below 10 MeV.\\

When a shell starts, the energy $E_{s}$ is very upper to the baryon
mass energy. We can define the ratio, $\eta$, between these two
energies which is
\begin{eqnarray*}
\eta = \frac{E_{s}}{(M_{b}/N_{s})\, c^{2}} = \frac{E}{M_{b}\, c^{2}}
\gg 1
\end{eqnarray*}
where $M_{b}$ is the total baryonic mass ejected.\\

In the observer frame, the shell thickness, $\Delta r$, is
supposed to remain constant and equal to $r_{0}$ until the broadening
radius $r_{b}$ (\citet{Goodman86}, \citet{Meszaros93}). Beyond this radius, 
defining the Lorentz factor, $\Gamma$, of the baryonic matter, the thickness 
becomes
\begin{eqnarray}
\Delta r \simeq (r/c)\, \Delta v \simeq r/2\Gamma^{2}
\label{shell}
\end{eqnarray}

In the same frame, we also define a radius, $r_{s}$, where the
kinetic energy of baryonic matter reaches its saturation value. At
this moment, the Lorentz factor $\Gamma=\Gamma_{max}$ is close to $\eta$.\\

In the co-moving frame, according to the Lorentz transformation, the
shell thickness is given by
\begin{eqnarray*}
\Delta R = \Gamma\, \Delta r\, .
\end{eqnarray*}

If we consider an adiabatic expansion of the shell,
conservation of entropy in the co-moving frame for a radius lower to
$r_{s}$ is given by
\begin{eqnarray}
r^{2} \, \Gamma \, r_{0} \, T^{3} = constant \, .
\label{ent}
\end{eqnarray}
At the same time, conservation of energy is such that
\begin{eqnarray}
r^{2} \, \Gamma^{2} \, r_{0} \, T^{4} = constant\, .
\label{enr}
\end{eqnarray}
  From (\ref{ent}) and (\ref{enr}), we deduce two laws of evolution
for $r\leq r_{s}$ which are $\Gamma(r) \propto r$ and $T(r)
\propto r^{-1}$. Then, the saturation radius $r_{s}$ can be defined by
\begin{eqnarray*}
r_{s}=\eta \, r_{0}\, .
\end{eqnarray*}
For $\eta$ of the order of 300, we obtain $r_{s}\simeq 3\times
10^{9}$ cm.\\

Beyond the radius $r_{s}$, the equation (\ref{ent}) is always valid
but the Lorentz factor of the shell remains constant. In this case,
the temperature is such that $T(r) \propto r^{-2/3}$. Considering
$\Delta r \geq r_{0}$, the radius $r_{b}$, according to
(\ref{shell}), satisfies $r_{b}\geq \Gamma^{2}\, r_{0}$. At
last, $\Gamma$ is supposed to reach its saturation value, $\eta$,
around $r_{s}$ well before $r_{b}$. In fact, $\eta$ is the average
value of the saturation bulk Lorentz factor, and we have to bear in mind that deviations from this average is expected to generate internal shocks. The radius
$r_{b}$ is so defined by
\begin{eqnarray*}
r_{b}\simeq \eta^{2}\, r_{0}\, .
\label{rb}
\end{eqnarray*}
For $\eta$ of the order of 300, we obtain $r_{b}\simeq 9\times
10^{11}$ cm.\\

To conclude, all the different parameters of a shell we need, can be
summarized by the next following expressions :\\

\hspace{-0.6cm}In the observer frame, the thickness of a shell is
\begin{eqnarray*}
\Delta r \simeq \left\{ \begin{array}{ccl}
& r_{0} & \hspace{0.8cm} {\rm for}  \hspace{0.3cm} r\leq r_{b} \\
& r/\Gamma_{max}^{2} \simeq r/\eta^{2}& \hspace{0.8cm} {\rm for}
\hspace{0.3cm} r\geq r_{b}
\end{array} \right.\, .
\end{eqnarray*}
The Lorentz factor of a shell is such that
\begin{eqnarray*}
\Gamma(r) \simeq \left\{ \begin{array}{ccl}
& r/r_{0} & \hspace{0.8cm} {\rm for}  \hspace{0.3cm} r\leq r_{s} \\
& \Gamma_{max}\simeq \eta & \hspace{0.8cm} {\rm for} \hspace{0.3cm}
r\geq r_{s}
\end{array} \right.\, .\\
\end{eqnarray*}

\hspace{-0.6cm}In the co-moving frame, the thickness of a shell
becomes equal to
\begin{eqnarray}
\Delta R \simeq \left\{ \begin{array}{ccl}
& \Gamma\, r_{0} \simeq r& \hspace{0.8cm} {\rm for}
\hspace{0.3cm}
r\leq r_{s} \\
& \Gamma_{max}\, r_{0} \simeq \eta \, r_{0}& \hspace{0.8cm}
{\rm for}  \hspace{0.3cm} r_{s}\leq r\leq r_{b} \\
& r/\Gamma_{max} \simeq r/\eta& \hspace{0.8cm} {\rm for}
\hspace{0.3cm}
r\geq r_{b}
\end{array} \right.\, .
\label{DELTAR}
\end{eqnarray}
The temperature of the pair electron-positron plasma is such that
\begin{eqnarray*}
T(r)\simeq \left\{ \begin{array}{ccl}
& 10 \times \left( \frac{T(r_{0})}{10\, MeV} \right) \left(
\frac{r_{0}}{r} \right) \, \, MeV & \hspace{0.8cm} {\rm for}
\hspace{0.3cm} r\leq r_{s} \\
& 10 \times \left( \frac{T(r_{0})}{10\, MeV} \right) \left(
\frac{r_{0}}{r_{s}} \right) \left( \frac{r_{s}}{r} \right)^{2/3}\, \,
MeV
& \hspace{0.8cm} {\rm for} \hspace{0.3cm} r\geq r_{s}
\end{array} \right.\, ,
\end{eqnarray*}
and the co-volume of a shell, $V_{c}(r)$, is
\begin{eqnarray}
V_{c}(r) \simeq \left\{ \begin{array}{ccl}
& \Omega \, r^{3} & \hspace{0.8cm} {\rm for}  \hspace{0.3cm}
r\leq
r_{s} \\
& \Omega\, \eta \, r_{0}\, r^{2}& \hspace{0.8cm} {\rm for}
\hspace{0.3cm} r_{s}\leq r\leq r_{b} \\
& (\Omega/\eta)\, r^{3} & \hspace{0.8cm} {\rm for} \hspace{0.3cm}
r\geq r_{b}
\end{array} \right.,
\label{COVOL}
\end{eqnarray}
where $\Omega$ ($\simeq 4\pi/500$) is the opening angle of the
emission. \\

The collision date (if any\ldots), $t_{c}$, between two shells of
Lorentz factor
$\gamma_{1}$ and $\gamma_{2}$, which started at two different times
separated by
$\Delta t_{0}$ is such that $t_{c} \simeq
2\frac{\gamma_{1}^{2}\gamma_{2}^{2}}{\vert
\gamma_{1}^{2}-\gamma_{2}^{2}\vert}
\Delta t_{0}$, which leads to collisions at various distances $r_{c}$
such that $r_{c}/r_{b} \sim \Delta t_{0}/t_{0}$ (see \citet{Daigne98}). For a long GRB
having a duration $t_{w} \sim 10$ s, the maximum collision radius
reaches the deceleration radius $r_{d}$. The light curve duration
could be due to either the duration of the wind $t_{w}$ or to the
time spread of the shell at the most remote collision, namely, $\Delta
t_{d} = t_{d}/\eta^{2}$. It turns out that this latter time is
comparable to the duration of the wind of a long GRB, namely,
$1-10$ s. The observed millisecond variations should come from
internal collisions located at a few $r_{b}$ where $\Delta r_{b} \sim
r_{0}$.

\hspace{-0.6cm}Anyway after a few $r_{b}$, all the shells mix up and
form a single jet. That is the reason why volume and thickness of a
shell can really be defined only before the broadening radius. After
this radius, in the equations (\ref{DELTAR}) and (\ref{COVOL}), they are undervalued.

\subsection{Conversion of the GRB energy into cosmic
rays}\label{subsec.}

During the evolution, the energy of the fireball is shared between
several forms : the thermal energy $E_{th}$, the magnetic energy
$E_{m}$, the bulk kinetic energy, the cosmic ray energy $E_{\star}$
and the energy in the form of hydromagnetic perturbations
$E_{m}^{\star}$. The energy share $E_{m}^{\star}$ is the reservoir for 
particle acceleration. \\

Until the saturation radius $r_{s}$, the ratio of the pair thermal
energy over the fireball energy, $(a\,T^4\,
V_{c})/E_{s}$, starts with a value close
to unity and then decays as $1/r$.
In this stage, the proton population is a tiny contribution both
in number of particles and in energy. Assuming that the protons
dominate the baryonic load, we get the total number of protons
\begin{equation*}
             N_{p} \simeq \frac{M_{b}}{m_{p}} \simeq 0.67 \times
10^{51}
             \left(\frac{\eta}{300}\right)^{-1}\left(\frac{E}{10^{51}
erg}\right)\,.
             \label{eq:NP}
\end{equation*}
Thus, the density of protons by shell, $n_{p}(r)$,  in the co-moving
frame and before the broadening radius, is such that
\begin{eqnarray*}
n_{p}(r) \geq 10^{-4}\left(\frac{t_{0}}{1\,ms}\right)
\left(\frac{t_{w}}{10\,s}\right)^{-1}\frac{N_{p}}{V_{c}(r)}\,,
\end{eqnarray*}
and the maximum value of $n_{p}(r)$ is equal to $N_{p}/V_{c}(r)$.\\
For $t_{0}=r_{0}/c=1$ ms, we can write
\begin{eqnarray*}
n_{p}(r) \geq \left\{ \begin{array}{ccl}
& 2.7\times 10^{27} \left(
\frac{t_{w}}{10\,s}\right)^{-1}\left(\frac{\eta}{300}\right)^{-1}
\left(\frac{\Omega/4\pi}{2\times
10^{-3}}\right)^{-1}\left(\frac{E}{10^{51}
erg}\right) \left(\frac{r}{r_{0}}\right)^{-3}  cm^{-3} \\
& 1.0\times 10^{20} \left( \frac{t_{w}}{10\,s}\right)^{-1}
\left(\frac{\eta}{300}\right)^{-2} \left(\frac{\Omega/4\pi}{2\times
10^{-3}}\right)^{-1}\left(\frac{E}{10^{51}
erg}\right) \left(\frac{r}{r_{s}}\right)^{-2}  cm^{-3}
\end{array} \right.
\end{eqnarray*}
respectively for $r\leq r_{s}$ and $r_{s}\leq r\leq r_{b}$.\\
Considering $\eta=300$, even if the GRB has a long duration about 10
s, the density of protons, at $r=r_{s}$, is larger than $10^{20}$
cm$^{-3}$.
We note that, up to the saturation radius, the
density of a shell is comparable to the density of a solid.\\

The corresponding ratio of the thermal energy of protons, $3
T/(2\eta\,
m_{p}c^{2})$, starts with a very low value ($\sim 10^{-4}$) and decays
like $1/r$ as well. The cosmic ray component is supposed to develope
out of the proton thermal component and would grow up to several
percents of $E$ to account for a significant contribution of the UHE cosmic
ray generation in the Universe. Assume that a fraction $\xi_{\star}$
of
the proton population is injected into the cosmic ray component (i.e.
$N_{\star} = \xi_{\star}N_{p}$) and that they reach a mean energy
$\bar \epsilon(r) = \bar \gamma(r)\, m_{p}\,c^{2}$ in the co-moving
frame. Then, the contribution to
the energy of  the fireball, in the observer frame, is given by
\begin{equation*}
             E_{\star}(r) = \frac{\xi_{\star} \bar
\gamma(r)\,\Gamma(r)}{\eta}\, E
             \label{ESTAR}
\end{equation*}
When the radius $r$ reaches the saturation radius $r_{s}$, the energy
of the cosmic ray component is equal to $\xi_{\star} \bar
\gamma(r)\,E/(1+4\xi_{\star}\bar \gamma(r)/3)$.
For a  cosmic ray spectrum in $\epsilon^{-2}$, $\bar \gamma = \log
(\gamma_{max})$, which clearly shows that the goal of converting
about $10$ percents of the fireball energy into
cosmic rays energy is achieved when a sizeable fraction
of the protons are injected in the cosmic ray population.

Regarding the magnetic energy of the fireball, two points of view can
be considered; either the magnetic field behaves like in a jet and
the poloidal component decreases like $1/r^{2}$ thus its pressure
decreases like the relativistic thermal pressure in $1/r^{4}$; whereas
the toroidal component decreases more slowly; or the magnetic field
of the shell disconnects from the central source; which is very
likely. In this latter case, the magnetic energy of the shell is
conserved as long as the dissipation is negligible (e.g. $B \propto
V_{c}^{-1/2}$) and decays when the Fermi processes and/or
reconnections become efficient in accelerating particles. Initially,
the intensity of magnetic field is usually considered as of the order
of
the equipartition value, which means that the magnetic energy is a
sizeable fraction of the fireball energy ($E_{m}(r_{0}) = \xi_{m} E$).

\subsection{The importance of pp-collisions during the primeval stage}

We begin with defining the radius $r_{\star}$ where a shell becomes
optically thin with respect to Compton scattering. It can easily be
checked that a typical shell width $\Delta R$ becomes smaller than the
flow transverse radius after a short while, when $r> \eta
\sqrt{\pi/4\Omega}\,r_{0}$ which is comparable to $r_{s}$. It will
turn out that the photosphere is located at a much larger distance for 
large enough $\eta$ (see (\ref{RST})) and
therefore the opacity of a shell is determined by its width.  The
Compton opacity is $\tau_{\star} = \sigma_{T}\, n_{e}\,\Delta
R$ where $\sigma_{T}$ is the Thompson cross section. We assume
$n_{e}\simeq n_{p} \simeq n_{b}$ and thus the co-moving baryon density
is related
to the wind mass flux $\dot M_{w}$ by
\begin{equation*}
        n_{b} = \frac{\dot M_{w}}{\Gamma\, \Omega\, r^{2}\, m_{p}\,c}
        \label{eq:NB}
\end{equation*}
Moreover for $r>r_{s}$, the GRB energy is converted into relativistic
kinetic
energy such that the kinetic luminosity (kinetic energy flux)
of the wind $L_{w} = E/t_{w} = \eta \,\dot M_{w}\,c^{2}$. For $r>r_{s}$, $\Gamma \simeq \eta$ and the Compton opacity is thus
\begin{equation*}
        \tau_{\star} = \frac{\sigma_{T}\,L_{w}}{\Omega
        \,m_{p}\,c^{3}\,\eta^{2}}\,\frac{\Delta R}{r^{2}} \, .
        \label{eq:TAUS}
\end{equation*}
We can define a critical value for $\eta$ such that the photospheric
radius is located at $r_{b}$, where shock acceleration starts. This
critical value $\eta_{\star}$ is given by
\begin{equation*}
        \eta_{\star} \equiv \left(\frac{\sigma_{T}L_{w}}{\Omega
        m_{p}c^{3}r_{0}}\right)^{1/5} \simeq 570 \times
\left(\frac{E}{10^{51}erg}\right)^{1/5}\left(\frac{t_{w}}{1s}\right)^{
-1/5}\left(\frac{\Omega/4\pi}{2\times 10^{-3}}\right)^{-1/5}
        \label{eq:ETAS}
\end{equation*}
The photospheric radius is such that
\begin{eqnarray}
r_{\star} = \left\{ \begin{array}{ccl}
r_{b} \,\left(\eta_{\star}/\eta\right)^{5} & \hspace{0.5cm} {\rm
for}  \hspace{0.3cm} \eta \leq \eta_{\star} \\
r_{b} \,\left(\eta_{\star}/\eta\right)^{5/2} & \hspace{0.5cm} {\rm
for}
\hspace{0.3cm} \eta \geq \eta_{\star}
\end{array} \right.\, .
\label{RST}
\end{eqnarray}
These simple formulae illustrate the requirement of a large value of
the asymptotic bulk Lorentz factor $\eta$ in order to observe an
optically thin X-ray spectrum. For $\eta$ larger than the critical
value, the internal shocks generate only a non-thermal spectrum just
after $r_{b}$, whereas, for $\eta < \eta_{\star}$, the internal
shocks start accelerating particles in an optically thick plasma.
Assuming that the GRB energy varies within two orders of magnitude,
the wind time by two orders of magnitude and the solid angle by one
order of magnitude, the possible values of $\eta_{\star}$ extend
within an interval of one order of magnitude. However it is worth
mentioning that this usual approach of the opacity issue is an
underestimate of the opacity effects, because the coalescence of
shells and ultimately their dissolution into a single jet increase the
opacity.\\

The possibility for the internal shock scenario to partially operate
in an optically thick regime is important when one consider
pp-collisions. Indeed, any relativistic protons can collide with other
protons (or neutrons) producing pions and thus neutrinos.
The cross section, $\sigma_{pp}$, for this production is constant and
equal to $2.7\times 10^{-26}$ cm$^{2}$ when the kinetic energy of
protons is larger than 1 GeV. The sheets become thin to
pp-collisions always before the photosphere since $\tau_{pp} =
n_{p}\,\sigma_{pp}\,\Delta R = 1$ at $r_{pp}$ such that $r_{pp}
=r_{\star}\sqrt{\sigma_{pp}/\sigma_{T}}<r_{\star}$ for $r_{pp}<r_{b}$
and $r_{pp}
=r_{\star}{\sigma_{pp}/\sigma_{T}}<r_{\star}$ for $r_{pp}>r_{b}$. In
this latter case, $r_{pp} \simeq 0.04\,r_{\star}$; which occurs when
$\eta \leq \eta_{\star}/2$. In this paper, we
intend to emphasize the importance of this energy limitation to proton
acceleration and estimate the resulting neutrino emission.

\section{Consequences of a fast acceleration regime}

In this section, we analyse the consequences of the usual assumption
of a Bohm scaling for the Fermi process for both the energy
distribution cut off and the depletion of the energy reservoir for
acceleration.

\subsection{Bohm scaling and energy losses}

We first consider a Fermi acceleration process that is assumed to
follow the so called Bohm's scaling namely characterized by its
time scale proportional to the Larmor time, i.e. :
\begin{equation*}
         t_{acc} = \kappa_{0}\, t_{L} = \kappa_{0}\, \frac{10^{-4}}{Z}
         \left(\frac{\epsilon}{1\,GeV}\right)\left(\frac{B}{1\,G}\right)^{-1}
\, s \ ,
         \label{eq:TACC}
\end{equation*}
where we take $\kappa_{0} =10$, as often assumed in astroparticle
physics.\\

In the early phase, the cut off of the proton energy distribution can
be caused by either the synchrotron loss or the pp-collisions; which
is obtained by equating the acceleration time and the loss time. The
synchrotron time $t_{syn}$ is given by
\begin{equation*}
             t_{syn}^{-1} =
\frac{4Z^{4}\sigma_{T}\,\gamma}{3m\,c}\,\left(
\frac{m_{e}}{m} \right)^{2}\,W_{m} ,
             \label{VM}
\end{equation*}
where $W_{m}$ is the density of magnetic energy. \\

\hspace{-0.6cm}This expression leads to the cut off energy :
\begin{equation*}
             \epsilon_{syn} \simeq 2.4\times 10^{11}
	    \frac{1}{\sqrt{\kappa_{0}}}
             \left(\frac{B}{1\,G}\right)^{-1/2} \,GeV \ .
             \label{VM}
\end{equation*}

We consider now the limitation caused by pp-collisions. The cut off
energy $\epsilon_{pp}$ is such that $(n_{p}\,\sigma_{pp}\,c)
\,\kappa_{0}\,t_{L} =1$, and thus
\begin{equation*}
             \epsilon_{pp} \simeq
	 \frac{10^4}{\kappa_{0}}
             \left(\frac{B}{1\,G}\right)
             \left(\frac{\eta^{6}}{\eta_{\star}^{5}}\right)
	 \left(\frac{r_{0}}{c}\right)
	 \, GeV \ .
             \label{VM}
\end{equation*}
At the broadening radius ($r_{b}$), where the magnetic field
could be as high as $10^{7}$ G, these two limitations are comparable :
\begin{equation*}
             \epsilon_{syn} \simeq 2.4\times 10^{7}
             \left(\frac{\kappa_{0}}{10}\right)^{-1/2}
             \left(\frac{B}{10^{7}\,G}\right)^{-1/2}\, GeV \ ,
             \label{VM}
\end{equation*}
and
\begin{equation*}
             \epsilon_{pp} \simeq 3\times 10^{9}
	 \left(\frac{\kappa_{0}}{10}\right)^{-1}
             \left(\frac{B}{10^{7}\,G}\right)
	 \left(\frac{\eta}{300}\right)
	 \left(\frac{\eta}{\eta_{*}}\right)^{5}\, GeV
             \label{VM}
\end{equation*}

During the expansion the ratio between these two limitations evolves
such that
\begin{equation*}
\frac{\epsilon_{pp}}{\epsilon_{syn}} \simeq 1.2\times 10^{3}\,
\left(\frac{\kappa_{0}}{10}\right)^{-1/2}
     \left(\frac{\eta}{300}\right)
	 \left(\frac{\eta}{\eta_{*}}\right)^{5}
\left(\frac{B}{10^{7}G}\right)^{3/2}
\left(\frac{r}{r_{b}}\right)^{2} \;
\label{eq:*}
\end{equation*}
which suggests that for any decrease of the magnetic field in
$r^{-\alpha}$ with $\alpha$ between 1 and 2, the most severe
limitation would be due to synchrotron losses. However we will show
in the next subsection that such efficient Fermi acceleration would
dissipate the magnetic energy very rapidly.

\subsection{The depletion of the energy reservoir for acceleration}

Such an efficient acceleration produces a strong depletion of the
energy reservoir for particle acceleration. The previous results
would make sense only if the depletion of the energy reservoir for
particle acceleration is slow. This is exactly the purpose of this
subsection.\\

Let $E_{m}^{\star}$ be the energy in the hydromagnetic perturbations
(magnetic and kinetic energy)
involved in the acceleration of particles. We assume that this is a
sizeable fraction $\xi_{m}^{\star}$ of the total magnetic energy and
that
even the large scale magnetic field in the shell is mostly
disconnected from the central source after a short while and thus is
rather tangle. Therefore we assume that the total magnetic energy
dissipates at the same rate than $E_{m}^{\star}$. A minimum
dissipation
is assumed by considering its energy loss by particle acceleration
only.
According to (\ref{eq:TACC}), the acceleration power is $Q^{+} =
\frac{N_{\star}}{\kappa_{0}t'_{L0}}m_{p}c^{2}$, where $t'_{L0}$ is the
Larmor time for a proton of 1 GeV in the co-moving frame. The minimum
depletion is then governed by $\dot E_{m}^{\star} = -Q^{+}$, which
reads :
\begin{equation}
         \dot E_{m} = - \frac{E}{t'_{\star}}\, \frac{\bar
B}{B_{\star}} \, ,
         \label{eq:DEPL}
\end{equation}
where we define a characteristic time $t'_{\star}$ at a given radius
$\tilde r$ where the magnetic field has an intensity $B_{\star}$,
such that
\begin{equation*}
         t'_{\star} \equiv (\eta\kappa_{0}\,
\xi_{m}^{\star}/\xi_{\star})\, t'_{L0}(B_{\star}) \ .
         \label{eq:TDEP}
\end{equation*}
Clearly, this time measures the rate of depletion at the considered
radius $\tilde r$ and must be compared to the co-moving dynamical
time
$t'(\tilde r) =  \tilde r/\Gamma c$. Let us make an estimate at
$r_{b}$, the late
stage actually; for $B = 10^{7}$ G and $\epsilon =1$ GeV,
$t'_{\star} \simeq 3\times 10^{-8}$ s, whereas $t'(r_{b}) \simeq
10^{-2}$ s. This perturbation burning out within a few nanoseconds is 
not realistic at all.
Before this stage, the
depletion time is much shorter than the fireball dynamical time by a
factor proportional to $(B\,r)^{-1}$ that goes at best like $r^{1/2}$
for $B
\propto r^{-3/2}$ corresponding to magnetic energy conservation and
the perturbations are burnt out almost immediately, their energy being
radiated by photons and neutrinos. The observation of these perturbations in the
light curve of gamma emissions suggests a much slower depletion; which
we will analyse in the next section.

\section{The interest of a progressive acceleration regime}

The Bohm scaling is often used in astroparticle physics, which
sometimes can provide some rough estimate of the high energy cut off.
But the true scaling \citep{Casse01}
 depends on the
turbulence spectrum. 
We will assume that the perturbations are
distributed according to the Kolmogorov law. This will significantly
change the performance of the Fermi process and the conclusions about
cosmic ray generation.

\subsection{Kolmogorov scaling and energy losses}

In the co-moving frame, the acceleration time depends on the speed 
$\beta_{\star}\,c$ of magnetic perturbations that scatter particles 
on both sides of the shocks. Considering the same 
average velocity, we have the relation
\begin{eqnarray*}
t_{acc}=\frac{t_{s}}{\beta_{\star}^{2}}\, ,
\end{eqnarray*}
where the scattering time, $t_{s}$, can be expressed like
\begin{eqnarray*}
t_{s}=(\eta_{b}\,\rho^{\beta-1}\,\omega_{L})^{-1}
\end{eqnarray*}
with $\eta_{b}=\frac{<\delta B^{2}>}{\bar{B}^{2}}$ and
$\rho=\frac{r_{L}}{l_{c}}\leq 1$ where $\l_{c}$ is the mean
coherence length of the magnetic perturbations whose spectrum is 
supposed to be a power law of index $\beta$. According to the Kolmogorov
theory, one can take $\beta=5/3$ and, with
$\kappa_{0}=1/(\beta_{\star}^{2}\,\eta_{b})$, we get
\begin{eqnarray*}
\kappa=\kappa_{0}\,\rho^{-2/3} \ .
\end{eqnarray*}

\hspace{-0.6cm}Considering $l_{c}\lesssim \Delta R$,
\begin{eqnarray*}
\rho=\frac{r_{L}}{\Delta R}\simeq\frac{\epsilon}{Z\,e\,B\,\Delta R\,c}
\end{eqnarray*}
and we have
\begin{eqnarray*}
\kappa\simeq 2.2\times 10^{-3}\,
\left(\frac{\kappa_{0}}{10}\right)\,
\left(\frac{\epsilon}{1\,GeV}\right)^{-2/3}
\left(\frac{B}{1\,G}\right)^{2/3}
\left(\frac{\Delta R}{1\,cm}\right)^{2/3}.
\end{eqnarray*}
The synchrotron limitation becomes
\begin{eqnarray}
\epsilon_{syn}\simeq 1.2\times 10^{19}\,
\left(\frac{\kappa_{0}}{10}\right)^{-3/4}\,
\left(\frac{B}{1\,G}\right)^{-5/4}
\left(\frac{\Delta R}{1\,cm}\right)^{-1/2}\,GeV.
\label{esyn}
\end{eqnarray}
For $r=r_{b}$ ($\simeq 10^{12}$ cm), $B=10^{7}$ G and
$\Omega=4\pi/500$,
$\epsilon_{syn}\simeq 3.8\times 10^{5}$ GeV.\\

\hspace{-0.6cm}As for the pp-collision limitation,
\begin{eqnarray*}
\epsilon_{pp}\simeq
\left(\frac{\kappa_{0}}{10}\right)^{-3}
\left(\frac{\eta}{300}\right)^{3}
\left(\frac{\eta}{\eta_{\star}}\right)^{15}\left(\frac{B}{10^{7}\,G}\right)
\left(\frac{\Delta R}{10^{12}\,cm}\right)^{-2}
\left(\frac{r}{r_{b}}\right)^{6}\,GeV.
\end{eqnarray*}
For $r\geq r_{b}$, $\Delta R = r/\eta$, defining $r_{b\star} \equiv
\eta_{\star}^{2}r_{0}$, we obtain
\begin{equation}
       \epsilon_{pp}\simeq 2.0\times
10^{5}\left(\frac{\kappa_{0}}{10}\right)^{-3}
       \left(\frac{B}{10^{7}\,G}\right)
       \left(\frac{\eta_{\star}}{570}\right)
       \left(\frac{\eta}{\eta_{\star}}\right)^{8}
       \left(\frac{r}{r_{b\star}}\right)^{4} \, GeV.
       \label{eq:EPPB}
\end{equation}
This energy increases up to a maximum value reached at $r_{pp}$ (when it's
larger than $r_{b}$), namely,
\begin{equation}
       \epsilon_{pp}\simeq 3.5 \times
       10^{2}\left(\frac{\kappa_{0}}{10}\right)^{-3}
       \left(\frac{B}{10^{7}\,G}\right)
       \left(\frac{\eta_{\star}}{570}\right)
       \left(\frac{\eta_{\star}}{\eta}\right)^{4}
       \, GeV.
       \label{eq:EPPR}
\end{equation}
As long as $r<r_{pp}$, the energy limitation is clearly due to
pp-collisions rather than synchrotron losses.
These results show, first, that it is not possible to accelerate
protons
beyond 100 GeV before the fireball becomes thin for protons at
$r_{pp}$, for $\eta> \eta_{\star}$; second, that, for $\eta
<\eta_{\star}/2$, the proton energy could increase above
100 GeV. For instance, if $\eta = \eta_{\star}/3$, $r_{\star} \simeq
240\, r_{b}$, $r_{pp} \simeq 10\, r_{b}$, $B$ weakens by a
factor $(r_{pp}/r_{b})^{-\alpha} \simeq 3.2 \times 10^{-2}$ with
$\alpha = 3/2$ and thus compensates the increase due to $\eta$.
Therefore the energy remains of order 100 GeV. We have to
examine how this regime changes the rate of perturbation energy
depletion. \\
In the opposite situation where the protons are not accelerated beyond GeV in the opaque stage, for $\eta>\eta_{\star}$ and/or $B<10^{5}$ G at $r_{b}$, nucleosynthesis is possible as shown by Lemoine \citep{Lemoine02}.

\subsection{The depletion of the energy reservoir for
acceleration}

Taking into account the variation of $\kappa$ in the acceleration
power, we define a new characteristic time $t'_{\star}$ which is
quite different from the last one :
\begin{equation*}
         t'_{\star} \equiv (\eta\kappa_{0}<\rho^{2/3}>^{-1}\,
         \xi_{m}^{\star}/\xi_{\star})\, t'_{L0}(B_{\star}) \ .
         \label{eq:TDEP}
\end{equation*}
For a power law distribution, not harder than $\epsilon^{-2}$,
$<\rho^{2/3}> \sim \rho_{0}^{2/3}$, $\rho_{0}$ being the rigidity for
$1$ GeV. For bell type distribution of standard deviation $\bar
\epsilon$, $<\rho^{2/3}> \sim <\rho(\bar \epsilon)^{2/3}>$.
Compared to the Bohm regime, the Kolmogorov regime increases this time
by a factor $\rho_{0}^{-2/3}$. At $r_{b}$ we get a more extended time
of the order of $0.1$ s, which is much more realistic than in the Bohm 
scaling case. If $r_{pp}$ is significantly larger,
the depletion behaviour is changed such that $t'_{\star}/t'$ decreases
instead of increasing like in Bohm regime. Indeed in Kolmogorov
regime, this ratio scales
like $B^{-1/3}r^{-5/9}$, which decreases when $B$ decreases less
slowly than $r^{-5/3}$. 

For a more detailed estimate, we integrated the differential equation 
(\ref{eq:DEPL}) which, after some algebra, leads to the following result :
\begin{equation*}
             \frac{E_{m}}{E_{m}(r_{b})} =
	    \left[1-\frac{E}{E_{m}(r_{b})}
            \frac{t'(r_{b})}{5t'_{\star}(r_{b})}
            \left(1-\left(\frac{r}{r_{b}}\right)^{-5/6}\right)
	   \right]^{6} \ .
             \label{DEM}
\end{equation*}
In fact, the perturbations are burnt out in a fraction $\delta t'$ of
$t'_{\star}(r_{b})$, which can be obtained directly from the
differential equation (\ref{eq:DEPL}) :
\begin{equation*}
             \delta t' = t'_{\star}(r_{b})\frac{E_{m}(r_{b})}{E} \ ,
             \label{DET}
\end{equation*}
which confirms the previous statement of a reasonable depletion time.

\subsection{The pp-neutrino emission}\label{subsec.NU}

In this subsection, we give an evaluation of the number of emitted pp-neutrinos and we calculate their energy spectrum.\\

The pp-collisions produce neutrinos after some reactions which are
\begin{eqnarray*}
p + p \longrightarrow & D + \pi^{+}  \\
& p + p + a (\pi^{+} + \pi^{-}) + b \pi^{0}  \\
& p + n + \pi^{+} + a (\pi^{+} + \pi^{-}) + b \pi^{0} \\
& 2n + 2\pi^{+} + a (\pi^{+} + \pi^{-}) + b \pi^{0}
\end{eqnarray*}
The production of mesons $\pi^{-}$ and $\pi^{+}$ gives neutrinos et
$\mu$-mesons through the decay reactions
\begin{eqnarray}
\pi^{-} \longrightarrow \mu^{-} + \overline\nu_{\mu} \nonumber\\
\pi^{+} \longrightarrow \mu^{+} + \nu_{\mu}
\label{Pi}
\end{eqnarray}
and other neutrinos are produced after
\begin{eqnarray}
\mu^{-} \longrightarrow e^{-} +\overline\nu_{e} + \nu_{\mu}
\nonumber\\
\mu^{+} \longrightarrow e^{+} + \nu_{e}  + \overline\nu_{\mu}\,\,.
\label{Mu}
\end{eqnarray}
Let's write energy of different particles
\begin{eqnarray*}
m_{\pi^{\pm}} & \simeq & 140 \,  MeV/c^{2} \\
m_{\mu} & \simeq & 105 \,  MeV/c^{2} \,\, .\\
m_{e^{\pm}} & \simeq & 0.5  \, MeV/c^{2}
\end{eqnarray*}
Thus, neutrinos coming from (\ref{Pi}) have a minimum energy equal to
\begin{eqnarray*}
\epsilon_{\nu_{\mu}} \simeq \frac{m_{\pi^{\pm}}^{2}-m_{\mu}^{2}}{2\,m_{\pi^{\pm}}}
\simeq 30\,  MeV
\end{eqnarray*}
and decay of $\mu$-mesons (\ref{Mu}) gives neutrinos having a minimum
energy which varies from 25 to 50 MeV.\\
Let $f_{\nu}(\epsilon_{\nu})$ be the energy distribution of
neutrinos,
normalised such that $\int
f_{\nu}(\epsilon_{\nu})\,d\epsilon_{\nu}=n_{\nu}$,
the number density of neutrinos; and $f_{\star}(\gamma)$ the Lorentz
factor
distribution of relativistic protons, normalised such that
$\int f_{\star}(\gamma)\,d\gamma=n_{\star}$, the number density of
relativistic protons. The kinetic equation for neutrinos can be
written
in a simplified way as follows :
\begin{equation}
     \frac{\partial}{\partial t}f_{\nu}+c\,\vec n.\nabla f_{\nu}=
     \xi_{\nu}\,\nu_{pp}\int  f_{\star}(\gamma)\,
     \delta(\epsilon_{\nu}-\gamma_{c}\epsilon_{0})\,d\gamma \ ,
     \label{eq:EENU}
\end{equation}
where $\epsilon_{0}$ is the average energy of the neutrinos generated
by the pp-collision in the center of mass frame (above estimated), 
$\xi_{\nu}$ the average number of produced neutrinos
at each collision and $\gamma_{c}$ the Lorentz factor of the
collision frame.
The energy of the emitted neutrinos remain close to the minimum
values
because the pions do not take all the energy available above the
threshold, but just a little excess above it, the remaining energy
being kept by the proton or the neutron. Therefore the number
$\xi_{\nu}$ is
just a few. The Lorentz factor $\gamma_{c} \sim \sqrt{\gamma}$,
where $\gamma$  is the Lorentz factor of the relativistic proton. So
we can easily deduce that, for $f_{\star} \propto \gamma^{-s}$, the
energy flux of neutrinos is in $\epsilon_{\nu}^{-2s+2}$, between
$\epsilon_{0}$ and $\epsilon_{0}\sqrt{\epsilon_{pp}/m_{p}c^{2}}$, in
the co-moving frame. For $s=2$, a neutrino spectrum in
$\epsilon_{\nu}^{-2}$ is expected between 5 GeV and 150 GeV,
typically,
for the observer frame seing the GRB flow coming front.

Assuming an isotropic neutrino emission in the co-moving frame, the
power emitted is
\begin{equation}
     \dot E'_{\nu} = \Omega \,\xi_{\nu}\, \epsilon_{0} \int
\nu_{pp}\,n_{\star}\, r^{2}\,d(\Delta R) \ ,
     \label{eq:ENU0}
\end{equation}
where the integral over the width extends down to the lower bound at
$r_{pp}$, corresponding to $\tau_{pp}=1$. Thus $E'_{\nu}= t'_{w}\,
\Omega\, \xi_{\nu}\, \epsilon_{0}\, c\,n_{\star}\, r_{pp}^{2}$. The
quantity
$t'_{w}\,\Omega\, c\,n_{\star}\,r_{pp}^{2}$ is the number of
relativistic
protons $\xi_{\star}N_{p}$ which have flown through the
"proto"-sphere
during the GRB event. Finally, we get a simple formula giving the
amount
of neutrino energy in the co-moving frame :
\begin{equation}
     E'_{\nu} =
     \xi_{\star}\,\xi_{\nu}\,\frac{\epsilon_{0}}{m_{p}c^{2}}\,
     \frac{E}{\eta} \ .
     \label{eq:ENUP}
\end{equation}
For an observer seing the GRB shells coming front, the neutrino
energy
emission is multiplied by the bulk Lorentz factor, thus
$E_{\nu} = \xi_{\star}\xi_{\nu}(\epsilon_{0}/m_{p}c^{2})E$
which reasonably leads to $E_{\nu} \sim 10^{-3}-10^{-2}E$.
The number of emitted neutrinos by a GRB is simply
$\xi_{\nu}N_{\star}$. A neutrino telescope of collecting
surface
$A$ can receive $N_{\nu}A/4\pi D^{2}$ from a GRB exploding at a
distance $D$. For example, a GRB occuring at 100 Mpc can provide
with $10^{5}$ neutrinos crossing a $km^{2}$-detector. However, the number of events recorded by the detector is obtained by multiplying this number by the detection probability which is quite low.\\

In the same process, $\pi^{0}$-decay occurs and generates gamma
photons. An excess of these photons could be seen in the gamma spectrum;
they superimpose a spectrum in $\epsilon_{\gamma}^{-2s+2}$ on the
synchrotron spectrum due to the electrons. Because we focused on proton acceleration and neutrino emission, we disregarded the interesting issue of the neutron component which can decouple under some conditions \citep{Derishev99}.

\section{The stage of UHE Cosmic Rays acceleration}

Because of their magnetic field intensity and the size of their wind, the
GRBs are considered as possible accelerators of UHE-cosmic rays 
(\citet{Vietri95}, \citet{Waxman95}) : for $r>r_{b}$, particles could reach the maximal energy beyond which they are no longer confined, $\epsilon_{max}=Z\,e\,B\,r$, namely 
\begin{eqnarray*}
\epsilon_{max}=7.7\times 10^{21}\left(\frac{B}{10^{7}\,G}\right)\left(\frac{r}{10^{12}\,cm}\right)\,eV,
\end{eqnarray*}
for a proton ($Z=1$).

\subsection{Proton energy in the transparent stage}

We examine the loss limitation suffered by the protons after crossing
the "proto"-sphere. After the equation (\ref{esyn}), we have already given 
the value of the maximum energy due to the synchrotron loss at $r_{b}$; 
thus, it scales like :
\begin{equation}
     \epsilon_{syn} \simeq 3.8\times 10^{5}
     \left(\frac{B(r_{b})}{10^{7}G}\right)^{-5/4}
     \left(\frac{r}{r_{b}}\right)^{(5\alpha -2)/4}\, GeV \ .
     \label{eq:ESYNP}
\end{equation}
It will turn out that this value is above the threshold of the
p$\gamma$-process, but is far below the range expected for getting
UHE-Cosmic Rays. For $\alpha = 3/2$, this range would be reached at a few
100 $r_{b}$.\\

\begin{figure}[!h]
\begin{center}
\includegraphics[totalheight=12cm,angle=-90]{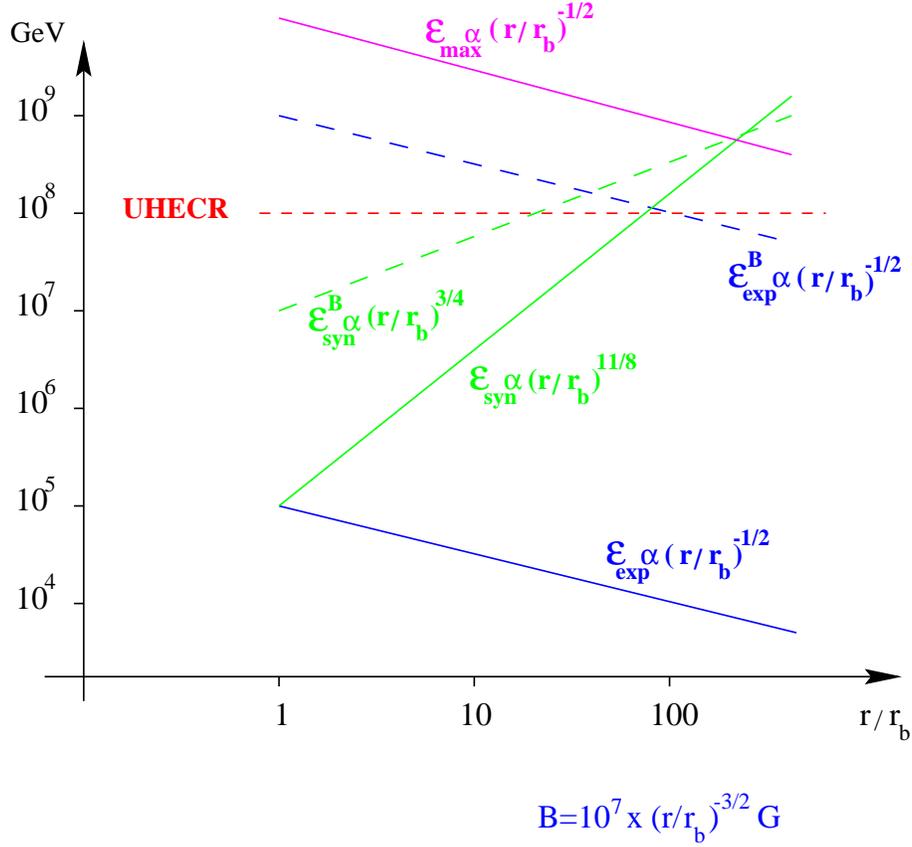}
\caption[10pts]{Diagram of energy limitation in the co-moving frame for $\alpha=3/2$ : the dashed lines show the result of the Bohm scaling and the solid lines are obtained with the Kolmogorov scaling. The horizontal dashed line represents the lowest UHECR range limit.}
\label{fig}
\end{center}
\end{figure}

The expansion loss can be significant. Indeed, the cut off energy is
obtained by setting $t_{acc} = t'=t/\Gamma$ in the co-moving frame, and the
limitation, $\epsilon_{exp}$, is such that
\begin{equation}
     \epsilon_{exp} \simeq 10^{5}\left(\frac{\kappa_{0}}{10}\right)^{-3}\left(\frac{\eta}{300}\right)
     \left(\frac{B(r_{b})}{10^{7}G}\right)
     \left(\frac{r}{r_{b}}\right)^{1-\alpha}\, GeV \ .
     \label{eq:EEXP}
\end{equation}
Whereas the synchrotron loss diminishes with distance, the
expansion limitation becomes more and more severe (see figure \ref{fig}). Moreover, it can
easily be checked that the escape time due to transverse diffusion is
always much longer than the expansion time.\\

\subsection{The p$\gamma$-neutrino emission}

A signature of VHE-protons acceleration is expected with the neutrino 
production resulting from the photo-production of pions, which is efficient to
produce neutrinos through the $\Delta-resonance$ (the so-called GZK effect) :
\begin{eqnarray}
              &  & \stackrel{(2/3)}{\rightarrow}  p + \pi^0
\rightarrow p
+
              \gamma + \gamma   \nonumber \\
             p + \gamma & \rightarrow  \Delta^+   &  \label{GZK} \\
              &   & \stackrel{(1/3)}{\rightarrow}  n + \pi^+
\rightarrow
...
             \rightarrow p + e^+ + e^- + \nu_e + \bar \nu_e +
\nu_{\mu} +
             \bar \nu_{\mu}  \nonumber
\end{eqnarray}
The threshold of the process is such that
$\epsilon_{p}\,\epsilon_{\gamma}\geq \frac{1}{2}m_{n}m_{\pi}
-\frac{1}{4}m_{\pi}^{2}$, which leads to a threshold for the proton
Lorentz factor $\gamma_{th} \sim 10^{4}$, the energy of the photons
being in the keV range in the co-moving frame. As previously seen,
$\gamma_{max}$ is at least ten times larger than $\gamma_{th}$.
The GRBs are likely opaque to this process until the shells reach the
radius $r_{p\gamma}$ such that
\begin{equation}
     r_{p\gamma} = \left(\frac{\sigma_{p\gamma}E_{\gamma}}{\Omega\,\bar
     \epsilon_{\gamma}}\right)^{1/2} \ ,
     \label{eq:RPG}
\end{equation}
where $\bar \epsilon_{\gamma}$ is the average energy of the target
photons in the observer frame (typically 1 MeV); which puts this
p$\gamma$-"proto"-sphere at a rather large radius of $1.5 \times
10^{15}$ cm for $E_{\gamma}=0.1\, E$.\\

In the collision frame, the energy of a pion, generated sufficiently
above the threshold, is $\epsilon'_{\pi} \simeq \epsilon'_{p}
\simeq \epsilon'_{\gamma} \simeq \gamma \epsilon_{\gamma}(1-\cos
\theta)$, where $\gamma$ is the Lorentz factor of the proton. The emitted
neutrinos have an energy which is a fraction $\alpha_{0}$
($\simeq 5\%$) of the pion energy.
Similarly to the statistical treatment of  the pp-collisions, we write
a simplified (with delta approximation instead of a function smoothed
by angle averaging) kinetic equation for the
isotropic distribution of the neutrinos generated by
p$\gamma$-collisions in the co-moving frame, which is :
\begin{equation}
             \frac{\partial}{\partial t}f_{\nu}+c\,\vec n.\nabla f_{\nu}=
     \xi_{\nu}\, c\,\sigma_{p\gamma}\int_{\gamma>\gamma_{th}}
     f_{\star}(\gamma)\, d\gamma \int 
     f_{\gamma}(\epsilon_{\gamma}) \,\delta(\epsilon_{\nu}-
     \alpha_{0}\gamma^{2} \epsilon_{\gamma})\, d\epsilon_{\gamma}\ .
             \label{KENUG}
\end{equation}
Integrating both sides of the equation over the co-volume and the
proper time, we easily get the number of emitted neutrinos :
\begin{equation}
     N_{\nu} =\xi_{\nu}N_{\star}^{>}=
     \xi_{\nu}N_{\star}\left(\frac{1}{\gamma_{th}}-
     \frac{1}{\gamma_{max}}\right) \ ,
     \label{eq:NUG}
\end{equation}
where $N_{\star}^{>}$ is the number of cosmic rays above the 
p$\gamma$-threshold energy, the last result being obtained for a
$\gamma^{-2}$ proton distribution. The integration, like in the case of
pp-collisions, starts at the p$\gamma$-"proto"-sphere.
The total energy radiated by neutrinos in the co-moving flow is also
derived easily from the kinetic equation :
\begin{equation}
     E'_{\nu} \sim
     \alpha_{0}\,\xi_{\nu}N_{*}\log\left(\frac{\gamma_{max}}{\gamma_{th}}\right)\,\bar \epsilon'_{\gamma} \ .
     \label{eq:EPNU}
\end{equation}
The logarithm factor is obtained for a
$\gamma^{-2}$ distribution of protons and can easily be modified.
This result can also be rewritten in the observer frame as follows :
\begin{equation}
     E_{\nu} \sim \alpha_{0} \,\gamma_{th}\,\bar \epsilon_{\gamma}
     \log\left(\frac{\gamma_{max}}{\gamma_{th}}\right) N_{\nu} \ .
     \label{eq:ENU}
\end{equation}
The energy radiated in the form of p$\gamma$-neutrinos is
$E_{\nu} \sim 10^{-7}E$. Numerical computation of such spectra has been done (see \citet{Mucke98}). The kinetic equation also provides with the
neutrino spectrum.
Assuming $f_{\star}$ is a power law distribution and that the energy
distribution of target photons is $\epsilon_{\gamma}\, f_{\gamma}
\propto \epsilon_{\gamma}^{-\alpha}$, we get a power law
energy spectrum for the neutrinos, namely
\begin{equation*}
\epsilon_{\nu}^{2}\,\frac{dN_{\nu}(>\epsilon_{\nu})}{d\epsilon_{\nu}} \propto
\epsilon_{\nu}^{-\frac{s-1}{2}} \ ,
\end{equation*}
in an energy range depending on the energy range
of the protons since $\epsilon_{\nu} \simeq \alpha_{0}\, \gamma^{2}
\epsilon_{\gamma}$. These spectra are given in the co-moving frame.
For the observer, they are Doppler beamed with the bulk Lorentz factor
of the relativistic wind. However, the number of events is so low that it is still ``virtual'' to talk about a spectrum...

\section{Discussion}

In  order to account for the non-thermal and highly variable gamma
emission of GRBs, the fireball model and the internal shock model have
been designed with baryonic load parameter $\eta$ that has been supposed
large enough to get a relativistic wind achieving a large
Lorentz factor $\Gamma \sim \eta$. However, it turns out that opacity
effects could easily be significant at the beginning of the emission.
We stress that point through a discussion involving a critical value
$\eta_{\star}$ of the baryon load parameter. 
Indeed one emphasis of
this paper is to analyse the opacity of the GRB to relativistic
protons with respect to pion production by pp-collisions. Actually, the
opacity condition relative to pp-collision is not far from the
Compton opacity condition. Therefore it is reasonable to think that a
significant fraction of the GRBs experience proton Fermi acceleration
with efficient pp-collisions revealed by neutrino radiation.

During a stage of pp-opacity, we have shown that Fermi
acceleration leads to
conclusions about the pp-process that are very sensitive to
the choice of the efficiency law of the acceleration. We have shown
that the Bohm scaling assumption leads to an efficient proton acceleration
that would be limited by synchroton loss, whereas the correct law
governed by the turbulence spectrum (Kolmogorov law was used) leads to
very different estimates and the pp-collisions process turns out to be
the main limitation of the proton energy. Moreover, the excessive
efficiency of the Bohm scaling would make the acceleration to
deplete its energy reservoir in a time too short to maintain the gamma
emission. The most important surprise raised in estimating the energy
limitation in the radiation free stage. Indeed, the expansion
limitation turns out to be drastic with the Kolmogorov law and
maintains the proton energy below a few $10^{5}$ GeV.

The analysis of the paper indicates that a double neutrino emission
can be expected with many GRBs, namely a stage of pp-neutrino
emission followed by a stage of p$\gamma$-neutrinos. The number of
emitted pp-neutrinos gives the amount of relativistic protons; and
the number of p$\gamma$-neutrinos gives the number of protons above
the threshold, which is about $10^{-4}$ less. We proposed an analytical
shape of the neutrinos spectra, as well as the photon spectrum generated
by the $\pi^{0}$-decay.\\
The magnetic field intensity is an important parameter that controls
the proton acceleration and the synchrotron losses. The neutrino emissions
significatively depend on its value at the crucial distance $r_{b}$.
We took a high but still reasonable value of $10^{7}G$ at this distance; if we  take
less, the synchrotron limitation is less important, but the expansion
limitation, which controls the highest  energy of the protons in the
GRB, becomes more severe, and also the pp-neutrino emission
becomes less energetic. 
If we unduly take more, synchrotron losses dominate
over pp-collisions and the acceleration is more efficient
against the expansion losses, however not sufficient to get
UHE-cosmic rays\ldots This paper does not exclude the possibility of
UHE-cosmic ray generation in GRBs. It simply states that its
achievement with Bohm scaling is not reliable and leads to
observational inconsistencies and that its achievement with
Kolmogorov scaling is impossible\ldots We think that there is
another possibility (\citet{Pelletier99}, \citet{PelletierK00}) 
that deserves a detailed investigation
that we will present in a forthcoming paper.\\

Acknowledgement :
The authors are grateful to Fr\'ed\'eric Daigne, Gilles Henri,
Martin Lemoine and Eli Waxman for fruitful discussions.

\bibliographystyle{aa}
\bibliography{/gagax1/ur1/dgialis/THESE/BIBFILES/biblio}

\end{document}